\documentclass[fdp,a4paper,fleqn%
]{w-art}
\usepackage{times,cite,w-thm}
\usepackage[]{graphicx}
\usepackage[latin1]{inputenc}
\usepackage[breaklinks=true,colorlinks=false,linkcolor=black,urlcolor=black,pagecolor=black]{hyperref}
\usepackage{amsfonts,amssymb,amsmath,amsthm,bbm}
\begin{document}
\DOIsuffix{theDOIsuffix}
\Volume{}
\Month{}
\Year{}
\pagespan{1}{}
\Receiveddate{XXXX}
\Reviseddate{XXXX}
\Accepteddate{XXXX}
\Dateposted{XXXX}



\title[Strange Metals and Dilatonic Black Holes]{Strange Metallic Behaviour and the Thermodynamics of Charged Dilatonic Black Holes}


\author[R.~Meyer]{René Meyer\inst{1,}%
  \footnote{Corresponding author and conference speaker, \quad E-mail:~\textsf{meyer@physics.uoc.gr},
            Phone: +30\,2810\,39\,4233, 
            Fax: +30\,2810\,39\,4274}}
\address[\inst{1}]{Physics Department, 
University of Crete, 
P.O. Box 2208, 
71003 Heraklion, Crete, Greece.}
\author[B.~Goutéraux]{Blaise Goutéraux\inst{2}}
\address[\inst{2}]{APC, Univ. Paris-Diderot, CNRS UMR 7164, F-75205 Paris Cedex 13, France.
}
\author[B.~S.~Kim]{Bom Soo Kim\inst{1,3}\footnote{R.M. and B.~S.~Kims work were partially supported by a European Union grant 
FP7-REGPOT-2008-1-CreteHEPCosmo-228644, Excellence grant MEXT-CT-2006-039047 and by ANR grant STR-COSMO, ANR-09-BLAN-0157.
}}
\address[\inst{3}]{IESL-FORTH, 71110 Heraklion, Crete, Greece}
\begin{abstract}
  We review a recent holographic analysis \cite{CGKKM} of charged black holes with scalar hair in view of their applications to the cuprate high temperature superconductors. We show in particular that these black holes show an interesting phase structure including critical behaviour at zero temperature or charge, describe both conductors and insulators (including holographic Mott-like insulators),  generically have no residual entropy and exhibit experimentally observed scaling relations between electronic entropy, specific heat and (linear) DC resistivity. Transport properties are discussed in the companion contribution \cite{BSKProc} in these proceedings.
\end{abstract}
\maketitle                   





\section{Introduction and Summary}

\begin{figure}\label{fig:phase}
\includegraphics[height=4.5cm]{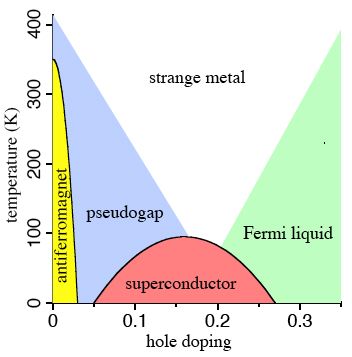}\hspace{0.5cm}
\includegraphics[height=4.5cm]{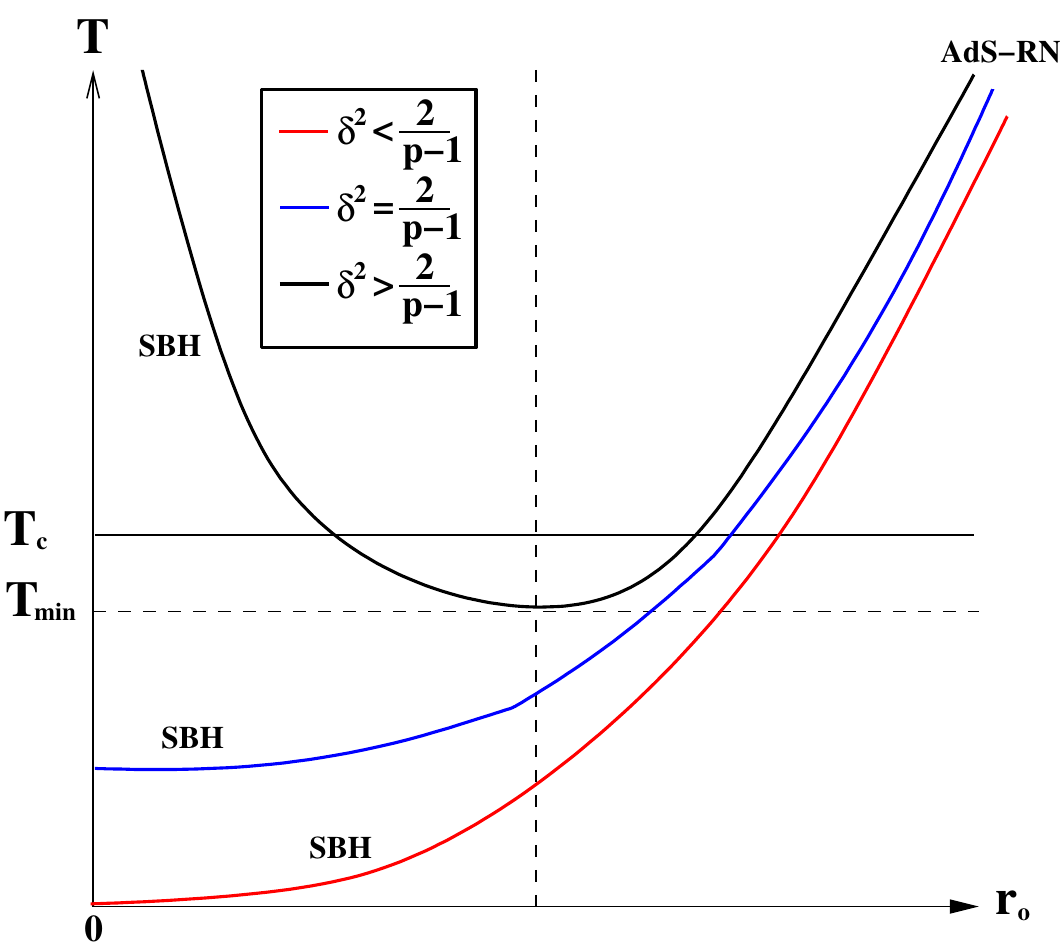}\hspace{0.5cm}
\includegraphics[height=4.5cm]{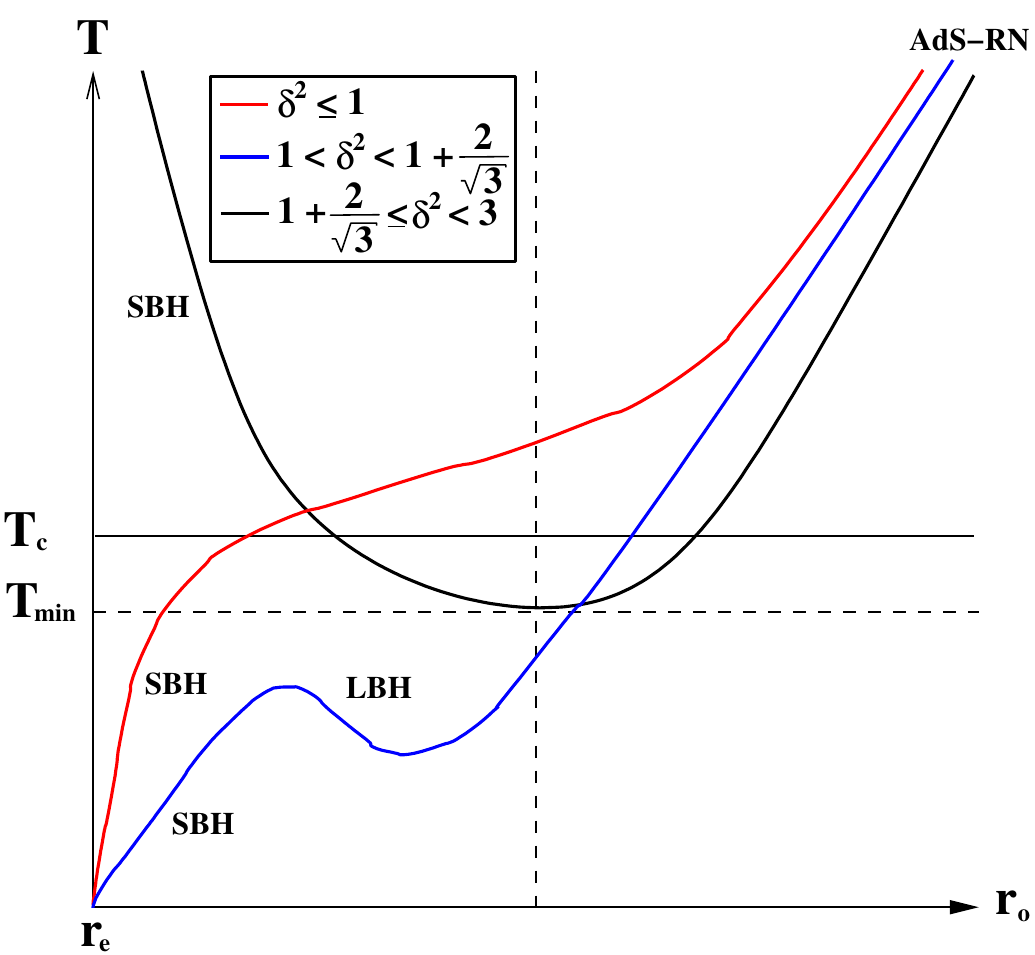}
\caption{\textbf{Left:} Phase diagram of hole-doped cuprates. In other systems the pseudogap region is much smaller, and the superconducting region can be nearly non-existing. \textbf{Middle:} Solution branches of the uncharged dilatonic black holes. \textbf{Right:} Solution branches for the $\gamma\delta=1$ exact solutions. The two branches for the $\gamma=\delta$ exact solutions, $\delta^2 \leq 1$ and $1 < \delta^2 < 3$, respectively are qualitatively similar to the red and blue branch in this case.}
\end{figure}

Several classes of real-world materials, in particular the cuprate-based high temperature superconductors, the heavy fermion systems and the newly-found iron-based high-$T_c$ compounds show thermodynamic and transport properties that cannot be described by Landau's Fermi liquid (LFL) theory of nearly free quasiparticle excitations around the Fermi surface, which in essence are dressed electrons. This non-Fermi liquid (NFL) behaviour hints to the presence of strong coupling and strong correlations, possibly making the system amenable to a holographic approach. In particular the high-$T_c$ cuprates show the following rather universal features in the phase diagram (see left panel in fig.~\ref{fig:phase}), as well as interesting thermodynamic and transport properties:
\begin{itemize}

\item The phase diagram, spanned by temperature and doping, shows several distinct regions: At low doping and intermediary temperatures it is an antiferromagnetic insulator, which changes into an insulating phase with a pseudogap at slightly higher doping. At even higher doping the pseudogap disappears as one enters the strange metal phase showing NFL behaviour. At very large doping the charge carriers again behave as a LFL. At very low temperatures a superconducting region forms roughly around the point where the pseudogap-strange metal and strange metal-Fermi liquid crossover lines meet.

\item The DC resistivity is linear in temperature, $\rho = \rho_0 + A T$,  within the strange metallic region in a large range of temperatures and even down to nearly absolute zero if the superconducting instability is supressed by magnetic fields or doping with ferromagnetic atoms \cite{supressing}. In a LFL, $\rho\sim T^2$ due to electron-electron scattering at very low temperatures only, hence both the unusual power of $T$ as well as the apparent stability of this behaviour over a large energy range is a mystery. Also the Hall conductivity in an magnetic field shows NFL behaviour, $\sigma_{xx}/\sigma_{xy}\sim T^2$ instead of $\sigma_{xx}/\sigma_{xy}\sim \rho \sim T$ for the Drude result \cite{TylerMackenzie}.

\item The AC conductivity for frequencies $T \ll \omega \ll \Lambda$ scales with an unusual power of $-2/3$, $\sigma_{xx}(\omega)\sim \omega^{-2/3}$ instead of the $\omega^{-1}$ Drude tail in a LFL \cite{ZaanenNature}, $\Lambda$ being a cutoff related to the chemical potential of the system. Such scaling laws can be reproduced by assuming a quantum critical point governing the finite temperature physics of the strange metal region \cite{ZaanenNature}. In the range $\omega < T$, the conductivity seems to follow Drude behaviour with a single relaxation time $\tau_r = \hbar A/(k_B T)$ \cite{ZaanenNature}, and $A$ being a numerical factor of order 1. This is different from $\tau \sim \hbar E_F/(k_BT)^2$ for electron-electron scattering in a LFL. The relaxation dynamics of the system thus seems to be governed by $\hbar$ alone, a pure quantum effect.

\item The electronic entropy and associated specific heat of a few cuprate compounds have been measured, starting from \cite{Loram}, with the result that it scales linearly with temperature in the strange metallic region at optimal or slight overdoping, $C\sim S\sim T$. This is the behaviour of a LFL at low temperatures which however seems to persist over a large temperature range, as well as together with genuine NFL behaviour, making it noteworthy by itself.\footnote{R.M. thanks Anton Rebhan for pointing this out to him.}

\end{itemize}

In order to holographically reproduce the above transport and thermodynamical properties, the holographic dual needs to include at least a metric dual to the energy-momentum tensor $T_{\mu\nu}$, and a U(1) gauge field dual to a conserved  current $J_\mu$. From the latter the AC and DC conductivities can be inferred from the retarded current-current correlator via a Kubo formula. The simplest system with this field content is Einstein-Maxwell theory with a cosmological constant
\begin{equation}\label{EMT}
S_{EM} = \frac{1}{16\pi G_N} \int d^{p+1}x \left[R - 2 \Lambda - \frac{1}{4} F_{\mu\nu}^2 \right]
\end{equation}
and has been extensively investigated in several contexts \cite{RN}. Its only static charged black brane solution, the AdS-Reissner-Nordström black brane, has however several shortcomings in describing real-world condensed matter systems. There is the existence of a finite ground state entropy density, hinting at an instability due to an enormous ground state degeneracy at large $N$. Also, while the DC resistivity for massive dilute carriers \cite{BSKProc} matches the experimentally observed temperature scalings in the cuprates, the AC  conductivity for the same carriers scales like $\sigma\sim (i\omega)^{-1}$ instead, matching Drude behaviour \cite{PHST}.\footnote{In \cite{HL2} the $O(N^0)$ contribution to the conductivity due to one-loop corrections of fermions in an AdS-RN black hole background was considered and shown to reproduce marginal Fermi liquid behaviour and in particular the $\sigma \sim \omega^{-2/3}$ law. This approach, being very interesting in its own right, can however also not reproduce both $\sigma \sim \omega^{-2/3}$ and Drude behaviour at low frequencies with the same choice of parameters.} Since many of the transport properties of the AdS-Reissner-Nordström metric are dictated by the $AdS_2\times \mathbbm{R}^2$ extremal near-horizon geometry describing a (from the 4D point of view) nonrelativistic conformal fixed point, a possible way to approach these shortcomings is to break the IR conformal symmetry by coupling the Einstein-Maxwell theory to an additional scalar field (in some cases the dilaton) with a runaway potential in the IR and study charged dilatonic black holes (CDBHs). The generic form of the scalar potential and coupling function in the IR as derived from string theory or supergravity will then be an exponential, and the (up to field redefinitions) unique action used in \cite{CGKKM} reads
\begin{equation}\label{EMDS}
S_{EMS} = \frac{1}{16\pi G_N} \int d^{p+1} x \left[R - \frac12 (\partial\phi)^2 - 2 \Lambda e^{-\delta\phi} - \frac{e^{\gamma\phi}}{4} F^2 \right]\,.
\end{equation}
Here $\gamma$ and $\delta$ are two parameters, $\Lambda = \frac{p(p-1)-2\delta^2}{2\ell^2}$ sets the IR length scale $\ell$. The scalar  is dual to an uncharged scalar operator ${\cal O}_{\phi}$ which, similar to the gluon condensate in QCD, breaks conformal invariance. This action cannot support asymptotically AdS black holes \cite{Wiltshire} and thus has to be viewed as an IR approximation to a full action admitting AdS asymptotics. It should however be noted that in particular the IR transport properties obtained  from the CDBHs might not change much in an AdS completed setup as long as the completion is sufficiently simple and does not introduce new length scales into the problem.\footnote{The IR and UV length scales will in general be related to each other once the scalar field value at the UV fixed point is fixed.} The phase structure might then still change due to the appearance of a large BH branch of Reissner-Nordström type, which (as will be seen below) is crucial in obtaining a complete picture of the physics even in the IR.

The topic of the work \cite{CGKKM} underlying these proceedings was to analyse several known classes of BHs in view of their applications to the cuprates and in general condensed matter physics. The main focus hence lies on planar ($p=3$) systems like the cuprates or bulk systems ($p=4$), but also linear systems ($p=2$) or quantum dots could be interesting ($p=1$) to study. Previous works include \cite{Gubser} which analysed an analytically solvable AdS-completed case with $(p,\gamma,\delta)=(4,\sqrt{2/3},-1/\sqrt{6})$ and noted that this resolves the zero temperature entropy problem, as well as another case with $(p,\gamma,\delta)=(3,1/\sqrt{3},1/\sqrt{3})$. \cite{Kachru,ChenPang} analysed near-horizon scaling solutions for $(p,\gamma,\delta)=(3,\gamma,0)$ constructed in \cite{Taylor,CGS} and found $\sigma_{xx}(\omega)\sim \omega^2$ independent of $\gamma$, for massless carriers. In the following sections we review our analysis of four classes of dilatonic BHs (uncharged solutions, exactly known charged solutions for $\gamma=\delta$ and $\gamma\delta=1$ and near-extremal scaling solutions for general $(p,\gamma,\delta)$) in view of their thermodynamic properties\footnote{We will focus on the canonical ensemble which is of more relevance to cuprate physics, since doping controls the charge carrier density rather than the chemical potential. There are other condensed matter systems such as graphene where it is rather the chemical potential that is dialed experimentally. The grand canonical ensemble is also discussed at length in \cite{CGKKM}.} and make some cross connections to the transport properties explained in more detail in \cite{BSKProc}. We will show in particular that these CDBHs show an interesting phase structure including critical behaviour at zero temperature or charge, describe both conductors and insulators (in particular holographic Mott-like insulators, i.e. insulators at a finite background charge density), generically (except for $\gamma=\delta$) show no residual entropy and exhibit experimentally observed scaling relations between electronic entropy, specific heat and (linear) DC resistivity.

\section{Uncharged Solutions}

The uncharged solutions of \eqref{EMDS} have been fully classified using superpotential methods in sec.~2 of \cite{CGKKM}. The only two solutions with a good singularity in the sense of Gubser\footnote{I.e. with a singularity which is shielded by a regular horizon at finite temperature.} \cite{GubserSingularity} at $r=0$ are \textit{small planar black holes} with metric given in eq.~(2.47)~of~\cite{CGKKM} and the corresponding \textit{thermal gas solution}. For an AdS completed potential there will also be a \textit{large asymptotically AdS-RN black hole} branch at large horizon radii (i.e. at high temperatures). For the small black hole the temperature-horizon radius ($r_0$) relation is\footnote{Henceforth we set the IR scale $\ell=1$ for simplicity. The exact dependencies can be found in \cite{CGKKM}.}
\begin{equation}\label{1}
 4 \pi T = \left( \frac{2p}{(p-1)\delta^2} -1\right) r_0^{\frac{2}{(p-1)\delta^2}-1}\,,
\end{equation}
leading to three qualitatively different solution branches, schematically depicted in the middle panel of fig.~\ref{fig:phase}: \textbf{(i)} For $\mathbf{\delta^2 < \frac{2}{p-1}}$, the temperature reaches zero at vanishing horizon radius $r_0=0$ and grows without limit at large $r_0$, smoothly joining onto the large BH branch at some point. At any finite $T$ the black hole phase is thermodynamically preferred, and a continuous phase transition with order depending on $\delta^2$ appears as $T\rightarrow 0_+$.\footnote{The free energy scales as $F \sim T^{\frac{6-(p-1)\delta^2}{2-(p-1)\delta^2}}$ at small temperatures, and hence the transition is $n^{th}$ order in the window given by $n-1 < \frac{6-(p-1)\delta^2}{2-(p-1)\delta^2} < n$. In particular $n\geq 4$.} \textbf{(ii)} In the marginal case $\mathbf{\delta^2=\frac{2}{p-1}}$ the temperature stays finite at $r_0=0$. At low temperatures the system is in the thermal gas phase.\footnote{With ``thermal background'', ``extremal background'' etc. we henceforth mean the extremal solution in Euclidean space with time compactified, i.e. the thermal gas background.} The order of the transition depends crucially on subleading terms in $V(\phi)$, ranging from mean-field and non-mean-field continuous to BKT type \cite{Gursoy1,Gursoy2}. \textbf{(iii)} For $\mathbf{\delta^2>\frac{2}{p-1}}$ the temperature decreases as $r_0$ increases, until at some point the AdS large BH branch (which is thermodynamically stable) takes over, and temperature starts to rise again. At a critical temperature $T_c$ there is a first-order phase transition to the thermal background. For $T<T_c$ the system behaves much like in the confined phase of holographic QCD, with a free energy of $O(1)$, and a discrete excitation spectrum. 

In particular the spectrum of gauge field excitations, which governs the structure of the current-current correlator's spectral function and hence its retarded Green's function and the associated conductivity (see \cite{CGKKM,BSKProc} for details), in this range (iii) generically\footnote{The exception is a planar system $p=3$ and sufficiently large dimension for ${\cal O}_\phi$: For $\Delta >1$ the charged excitation spectrum is continuous and gapless, and for $\Delta=1$ continuous and gapped.} shows a discrete and gapped spectrum, while it is continuous and gapped for case (ii) and continuous and gapless for range (i). The system thus is a \textbf{conductor} in cases (i) and (ii), while it turns out to be an \textbf{insulator} in region (iii).\footnote{We define an insulator to have nonvanishing conductivity at small frequencies $\omega \rightarrow 0$. The system can be an insulator even if the charged excitation spectrum is continuous and gapless, since the behaviour of the effective potential for the gauge field excitations at the horizon is all that matters to determine this property: Either a wave can tunnel through the potential into the BH horizon, or it gets fully reflected by the potential. In the former case the retarded current-current Green's function (from which the conductivity is calculated via the usual Kubo formula $\sigma_{xx}(\omega) = G_{R,xx}(\omega,\vec{k}=0)/(i\omega)$) is finite, while in the latter case it vanishes. The first case thus is conducting, the latter insulating. The spectrum of charged excitations on the other hand also depends on the properties of the effective potential at infinity. For more details, see \cite{CGKKM,BSKProc}.} One should note that the value of $\delta^2$ cannot grow too large, since it is constrained to $\delta^2\leq 2(p+2)/3(p-1)$ above which the graviton fluctuation problem in the uncharged background seizes to be a well-defined Sturm-Liouville problem, rendering the problem holographically meaningless. 

\section{Exact Charged Solutions}

Two classes of charged black brane solutions of \eqref{EMDS} can be found analytically in 3+1 dimensions \cite{CGS}: The case $\gamma\delta=1$ (see eq.~(6.1) of \cite{CGKKM}) encompasses the AdS-Schwarzschild solution via the limit $\delta \rightarrow 0$, $q \sim \delta^{a>1} \rightarrow 0$. This case incorporates both Kaluza-Klein reduction of $\int R - 2\Lambda$ in 5D ($\gamma=\delta^{-1} = \sqrt{3}$), as well as 4D string theory with the vector arising from the NS sector ($\gamma=\delta=1$). 
The case $\gamma=\delta$ is also exactly solvable in $p=3$ (see eq.~(7.1) of \cite{CGKKM}, known since the second reference in \cite{Taylor}), leading to a qualitatively different class of rather Reissner-Nordström-like black holes, with the AdS-RN limit being $\gamma=\delta=0$. Both classes coincide for $\gamma=\delta=1$.

\paragraph{The case $\gamma\delta=1$:} The temperature-radius relation 
\begin{equation}\label{Tvsrgd1}
4\pi T = (3-\delta^2)r_+^{1-\delta^2}\left[ 1 - \frac{4(1+\delta^2) Q^2}{\delta^2(3-\delta^2)^2 r_+^{2(3-\delta^2)}} \right]^{1-\frac{2(\delta^2-1)^2}{(1+\delta^2)(3-\delta^2)}}
\end{equation}
yields three qualitatively distinct cases (see the right panel of fig.~\ref{fig:phase}): \textbf{(i)} For $\delta^2 \leq 1$ both powers in \eqref{Tvsrgd1} are nonnegative, leading to a single stable black hole branch with $T = 0$ at extremality which is always thermodynamically preferred. \textbf{(ii)} For $1 < \delta^2 < 1 + \frac{2}{\sqrt{3}}$ there is a small black hole branch with $T=0$ at extremality, a large BH branch with diverging $T$, and the large AdS-RN black holes with $T \sim r_+$ after AdS completion. Since both the small BH and AdS-RN branches are stable while the large BH branch is not, this case will show first-order phase transitions between the small and AdS-RN BHs which may be tunable to  second-order transitions for specific AdS-completed potentials. \textbf{(iii)} $1 + \frac{2}{\sqrt{3}} \leq \delta^2 < 3$ This range is similar to case (iii) of the uncharged solutions, with a single, unstable small BH branch with diverging $T$ at extremality, the AdS-RN black holes at larger $r_+$ being the stable high $T$ phase, and a first-order phase transition to the confining but charged extremal black hole background at low temperatures. Investigation of the transport properties \cite{BSKProc} shows that these backgrounds are holographic insulators for $1 + \frac{2}{\sqrt{3}} \leq \delta^2 < \frac{5+\sqrt{33}}{4}$, despite the existing charge density, i.e. they are \textbf{holographic Mott-like insulators}.

All $\gamma\delta=1$ solutions have vanishing entropy at extremality. The phase diagrams for (i) and (ii) are depicted in figs.~18~of~\cite{CGKKM} and have interesting boundaries: There is no critical behaviour for vanishing charge at finite $T$, but for vanishing $T$ at finite charge we find a $2^{nd}$ order transition to the extremal background for $1-\frac{2}{\sqrt{5}} < \delta^2 < 1 + \frac{2}{\sqrt{5}}$ which becomes $3^{rd}$ order outside that region. Exactly at $\delta^2 = 1 \pm \frac{2}{\sqrt{5}}$ there is no critical behaviour, but this could be changed by subleading terms in the potential similar to \cite{Gursoy1,Gursoy2}. These two points also show a DC resistivity for massive dilute charge carriers \cite{BSKProc} linear in $T$, $\rho = A T + \dots$.\footnote{This assumes that the scalar is not the dilaton, i.e. does not enter into the transition between string and Einstein frame.} For the planar case  $p=3$, the entropy and specific heat also scale in the same way,
\begin{equation}\label{scalings}
S \sim C_Q \sim \rho \sim T\,. 
\end{equation}
It is an interesting question whether the appearance of linear resistivity, entropy and specific heat scaling exactly at the border between second and third order critical behaviour in this system has any deeper physical meaning. 

\paragraph{The case $\gamma=\delta$:} Entropy, specific heat and DC resistivity (in the limit of very massive and dilute charge carriers and for $\phi$ not being the dilaton) of these BHs scale at low $T$ as
\begin{equation}\label{g=dproperties}\hspace{-0.5cm}
S = r_+^2 = r_{e}^2 + \frac{T}{2} \frac{r_{e}^{1+\delta^2}}{3-\delta^2} + \dots\,,\quad C_Q = \frac{T}{2} \frac{r_{e}^{1+\delta^2}}{3-\delta^2} + \dots\,,\quad \rho = \frac{T_f}{J^t} r_{e}^2 \left( 1 + \frac{2\pi r_{e}^{\delta^2-1}}{3-\delta^2} T + \dots \right)\,,
\end{equation}
with $T_f$ being the fundamental string tension, $J^t$ the charge density and the extremal horizon radius $r_{e}^4 = \frac{q^2}{4(3-\delta^2)}$. The residual entropy and the nonzero residual resistivity $\rho(T=0)$ are connected to each other, since both $S$ and $\rho$ are in this case determined by the xx-component of the metric in Einstein frame $g_{xx}^E(r_h)$, evaluated at the horizon. 

There are only two kinds of solution branches, analoguous to cases (i) and (ii) of the $\gamma\delta=1$ branches: \textbf{(i)} a single stable small BH branch for $\delta^2 \leq 1$ and \textbf{(ii)} a stable small and an unstable large BH branch (plus stable large AdS-RN BHs if the potential is AdS completed) for $1 < \delta^2 < 3$.\footnote{For a plot of the temperature-radius relation (without the large AdS-RN BHs) see fig.~(32) of \cite{CGKKM} which we do not reproduce here since it is so similar to the right panel of fig.~\ref{fig:phase}.} There are no Mott-insulating states in this class of solutions. The phase diagram in the canonical ensemble, depicted in fig.~(38) of \cite{CGKKM}, again shows interesting behaviour as the axes are approached: Contrary to the $\gamma\delta=1$ case there is now no critical behaviour as $Q>0, T\rightarrow 0$, the system smoothly goes to the extremal black hole background.  Approaching the zero charge axis, $T>0, Q\rightarrow 0$, the system however now undergoes a $2^{nd}$ order transition to the neutral black holes for $\delta^2 < 1$ and a $1^{st}$ order transition to the extremal background for $\delta^2>1$. Between these two behaviours, i.e. for $\delta^2=1$, no obvious observation concerning transport properties can be made this time, except probably that the $O(T^0)$ and $O(T^1)$ coefficients of both the entropy and the DC resistivity scale in the same way with $r_e^2 \sim q$. In this case $\rho \sim \frac{q}{J_t}(1 + A T + \dots)$ is proportional to the ratio between the background charge density $q$ and the density of massive charge carriers, which are dilute, i.e. $J_t \ll q$.

\section{Near-Extremal Scaling Solutions}

\begin{figure}\label{fig:SS}\hspace{-1.5cm}
\includegraphics[height=4.5cm]{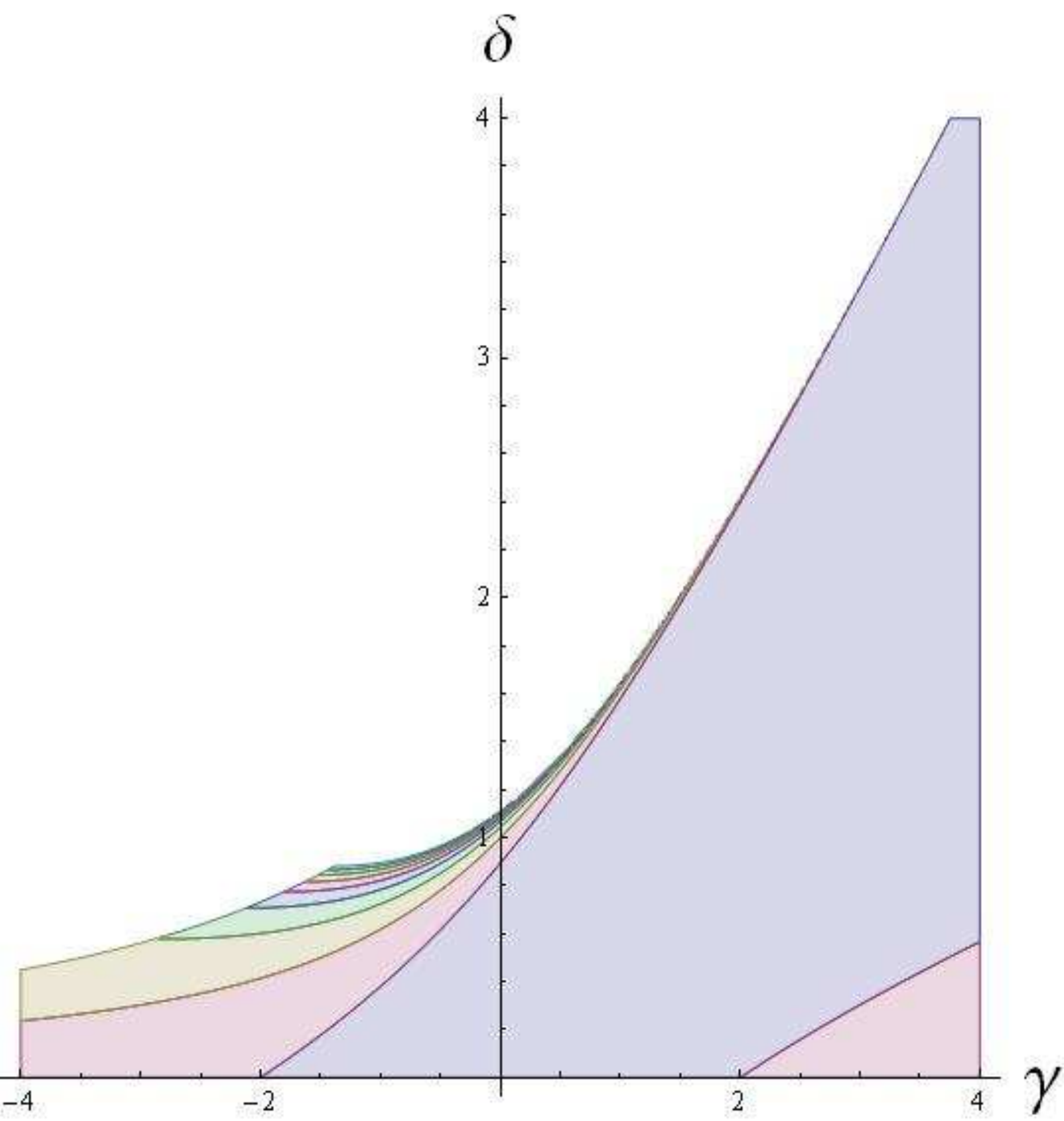}\hspace{0.5cm}
\includegraphics[height=4.5cm]{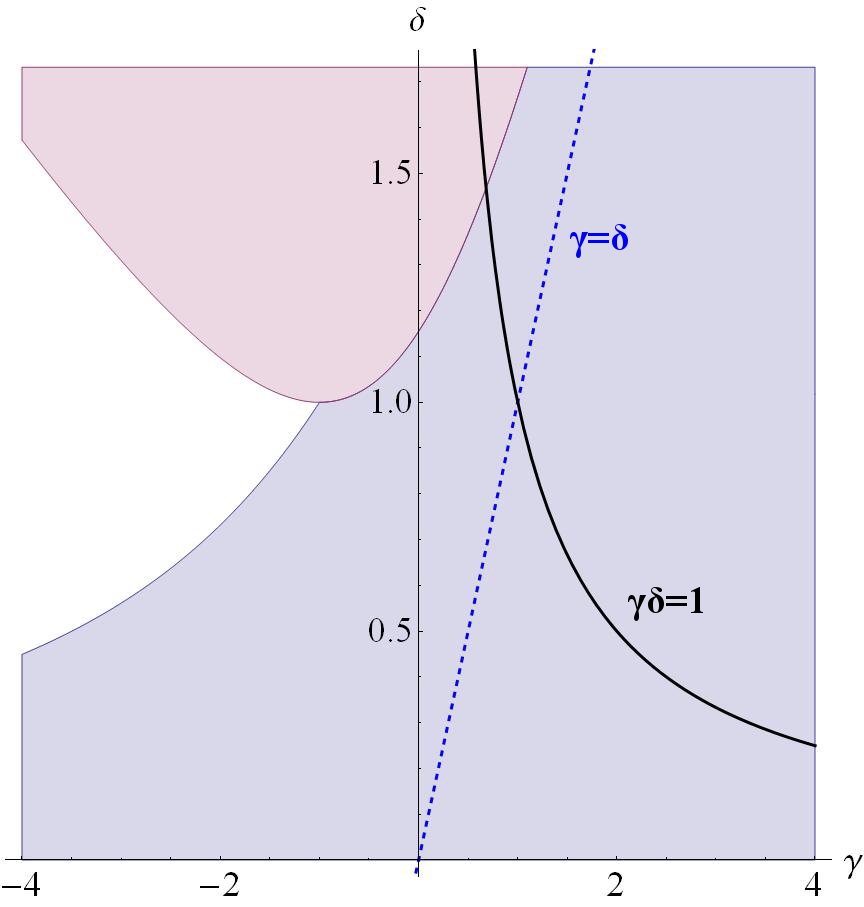}
\includegraphics[height=6cm]{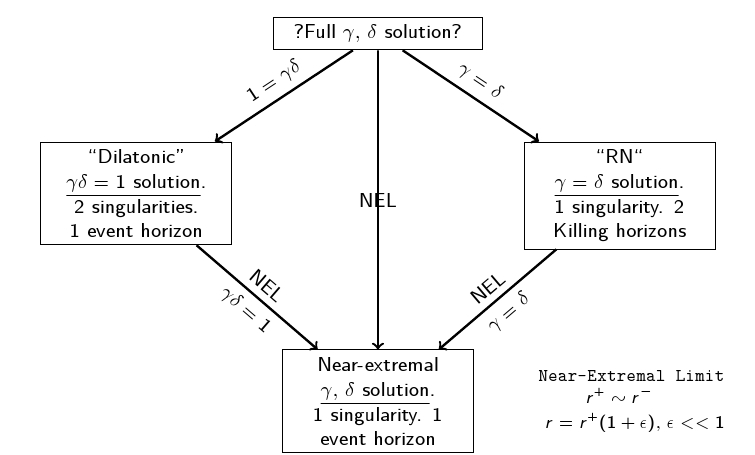}
\caption{\textbf{Left:} Regions of critical behaviour as $T\rightarrow 0$ for $p=3$, being $2^{nd}$ order in the blue region and $3^{rd}$ order in the purple region adjacent to the blue one. The stripes starting with yellow to
the left of the blue and purple regions are of $4^{th}$ (yellow) up to $10^{th}$ order. 
Above them even higher orders also occur, in smaller and smaller regions. \textbf{Middle:} Constraints on parameter space for $p=3$. The blue region is allowed by Gubser's constraint. In the purple region the temperature diverges at extremality. \textbf{Right:} The near-extremal scaling solutions are the near-extremal limits (NEL) of the exact charged solutions, which are known for $\gamma\delta = 1$ and $\gamma=\delta$, but unknown for general $(\gamma,\delta)$.}
\end{figure}

The only charged BH solutions of \eqref{EMDS} known currently for general $(\gamma,\delta)$ are near-extremal scaling solutions, obtained from a scaling Ansatz for the metric. This solution, given for general dimension $p$ in eqs.~(8.8)-(8.12)~of~\cite{CGKKM}, is not a general solution of Einstein's equations, as it lacks a tunable electric charge as the second integration constant. Instead, the charge is a function of $(\gamma,\delta)$ and $\Lambda$. These solutions thus can at best be approximations to general solutions, and indeed they are the \textbf{near-extremal approximations} to the $\gamma\delta=1$ and $\gamma=\delta$ class (see the right panel of fig.~\ref{fig:SS}). We suspect that this relation holds true as well for the yet unknown general solution for general $(\gamma,\delta)$ (it does for all known solutions we checked). The scaling solutions have been found independently in \cite{Cquest} and reduces  for $\delta=0$, $z = 1 + \frac{4}{\gamma^2}$ to the Lifshitz black holes of \cite{Taylor} with dynamical critical exponent $z$. The entropy and temperature of the scaling solutions are
\begin{equation}\label{SSentropy}
S \sim (2m)^{\frac{(p-1)(\gamma-\delta)^2}{w_pu}}\,,\quad T \sim (2m)^{1 - \frac{(p-1)(\gamma-\delta)^2}{w_pu}}\,,\quad w_pu = 2(p-1) + p \gamma^2 - 2\gamma\delta - (p-2)\delta^2\,,
\end{equation}
with $m$ being the nonextremality parameter in the solution ($m=0$ is the extremal limit). We thus find that there is \textbf{no residual entropy at extremality generically, with the exception of the case $\gamma=\delta$}. The temperature of the scaling solutions diverges in the region $\frac{(p-1)(\gamma-\delta)^2}{w_pu} > 1$ as extremality is approached, as happens in the $\gamma\delta=1$ solutions (the black curve in the right panel of fig.~\ref{fig:phase}). We expect these cases to include \textbf{holographic Mott-like insulators}, as discussed above for $\gamma\delta=1$, with a first-order phase transition to the extremal background at a critical temperature. The opposite case $\frac{(p-1)(\gamma-\delta)^2}{w_pu} \leq 1$ shows a \textbf{multitude of critical behaviours of any order} as $T=0$ is approached (see the right panel of fig.~\ref{fig:SS}).\footnote{The critical behaviour is of order $n$ in the range $n-1 < \frac{w_pu}{w_pu- (p-1)(\gamma-\delta)^2} < n$.} Gubser's constraint $w_pu > 0$, $u = \gamma^2-\gamma\delta + 2>0$, $v = \delta^2-\gamma\delta -2 < 0$ is the main restriction on $(\gamma,\delta)$ from the thermodynamic point of view (see the middle panel of fig.~\ref{fig:SS}). Analysing the graviton and gauge field fluctuation equations will further constrain parameter space, as discussed in the contribution \cite{BSKProc} to these proceedings.

%

\begin{thebibliography}{[1]}

\bibitem{CGKKM}
 C.~Charmousis, B.~Goutéraux, B.~S.~Kim, E.~Kiritsis and R.~Meyer, 
  JHEP {\bf 1011} (2010) 151. 

\bibitem{BSKProc}
  B.~Goutéraux, B.~S.~Kim and R.~Meyer, `'Charged Dilatonic Black Holes and their Transport Properties'', in ``Proceedings of the XVIth European Workshop on String Theory''.

\bibitem{supressing}
 S.~H.~Naquib~et.~al., Physica C {\bf 387}, 365 (2003); R.~Daou~et.~al., Nature Physics \textbf{5}, 31 (2009); R.~A.~Cooper~et.~al., Science {\bf 323}, 603 (2009); N. Doiron-Leyraud~et.~al., [arXiv: 0905.0964].

\bibitem{ZaanenNature}
  D.~van~der~Marel~et.~al.,
Nature\ {\bf 425} (2003) 271-274.

\bibitem{TylerMackenzie}
 A.~W.~Tyler~and~A.~P.~Mackenzie, Physica C {\bf 1185} 282-287 (1997).

\bibitem{Loram}
 J.~W.~Loram~et.~al., Phys.~Rev.~Lett. {\bf 71} 1740-1743 (1993).

\bibitem{RN}
  S.~A.~Hartnoll and C.~P.~Herzog,
  Phys.\ Rev.\  D {\bf 76} (2007) 106012 and 
  Phys.\ Rev.\  D {\bf 77} (2008) 106009; 
  S.~A.~Hartnoll, C.~P.~Herzog and G.~T.~Horowitz,
  Phys.\ Rev.\ Lett.\  {\bf 101} (2008) 031601 and  
  JHEP {\bf 0812} (2008) 015;
  H.~Liu, J.~McGreevy and D.~Vegh,
  arXiv:0903.2477 [hep-th];
  N.~Iqbal and H.~Liu,
  Fortsch.\ Phys.\  {\bf 57}, 367 (2009); 
  T.~Faulkner, H.~Liu, J.~McGreevy and D.~Vegh,
  arXiv:0907.2694 [hep-th].
  N.~Iqbal, H.~Liu, M.~Mezei and Q.~Si,
  arXiv:1003.0010 [hep-th].
  M.~Cubrovic~et.~al., 
  Science {\bf 325} (2009) 439;
  


\bibitem{PHST}
  S.~A.~Hartnoll, J.~Polchinski, E.~Silverstein and D.~Tong,
  JHEP {\bf 1004} (2010) 120.

\bibitem{HL2}
  T.~Faulkner, N.~Iqbal, H.~Liu, J.~McGreevy and D.~Vegh,
  Science {\bf 329}, 1043 (2010).




\bibitem{Wiltshire}
  S.~J.~Poletti and D.~L.~Wiltshire,
  Phys.\ Rev.\  D {\bf 50} (1994) 7260
  [Erratum-ibid.\  D {\bf 52} (1995) 3753].

\bibitem{Gubser}
  S.~S.~Gubser and F.~D.~Rocha,
  Phys.\ Rev.\  D {\bf 81} (2010) 046001.

\bibitem{Kachru}
  K.~Goldstein, S.~Kachru, S.~Prakash and S.~P.~Trivedi,
  JHEP {\bf 1008}, 078 (2010).

\bibitem{Taylor}
  R.~G.~Cai, J.~Y.~Ji and K.~S.~Soh,
  Phys.\ Rev.\  D {\bf 57} (1998) 6547; 
  M.~Taylor,
  [arXiv:0812.0530 [hep-th]].

\bibitem{ChenPang}
  C.~M.~Chen and D.~W.~Pang,
  JHEP {\bf 1006} (2010) 093.

\bibitem{GubserSingularity}
  S.~S.~Gubser,
  Adv.\ Theor.\ Math.\ Phys.\  {\bf 4} (2000) 679.

\bibitem{Gursoy1}
  U.~Gursoy,
  JHEP {\bf 1101} (2011) 086.

\bibitem{Gursoy2}
  U.~Gursoy,
  JHEP {\bf 1012} (2010) 062.

\bibitem{CGS}
 C.~Charmousis, B.~Gouteraux and J.~Soda,
 Phys.\ Rev.\  D {\bf 80} (2009) 024028.

\bibitem{Cquest}
B.~H.~Lee, S.~Nam, D.~W.~Pang and C.~Park,
  [arXiv:1006.0779 [hep-th]].



\end{thebibliography}
%

\end{document}